\documentclass[apj]{emulateapj}
\newcommand  \kms      {\ifmmode {\rm km\,s}^{-1} \else km\,s$^{-1}$\fi}

\newcommand  \cmii     {\hbox{cm$^{-2}$}}
\newcommand  \ergs     {\ifmmode {\rm ergs\,s}^{-1} \else ergs s$^{-1}$\fi}
\newcommand  \ergcms   {\ifmmode {\rm ergs\,cm}^{-2}\,{\rm s}^{-1}
                        \else ergs\,cm$^{-2}$\,s$^{-1}$\fi}
\newcommand  \ergcmsA {\ifmmode{\rm ergs\,cm}^{-2}\,{\rm s}^{-1}\,{\rm\AA}^{-1}
                        \else ergs\,cm$^{-2}$\,s$^{-1}$\,\AA$^{-1}$\fi}
\newcommand \ergcmsHz {\ifmmode{\rm ergs\,cm}^{-2}\,{\rm s}^{-1}\,{\rm Hz}^{-1}
                        \else ergs\,cm$^{-2}$\,s$^{-1}$\,Hz$^{-1}$\fi}
\newcommand  \phcms    {\ifmmode {\rm ph\,cm}^{-2}\,{\rm s}^{-1}
                        \else ,ph\,cm$^{-2}$\,s$^{-1}$\fi}
\newcommand  \phcmsA   {\ifmmode {\rm ph\,cm}^{-2}\,{\rm s}^{-1}\,{\rm\AA}^{-1}
                        \else ph\,cm$^{-2}$\,s$^{-1}$\,\AA$^{-1}$\fi}

\newcommand{\MBH}{$M_{\rm BH}$}

\def\micron{\ifmmode \mu{\rm m} \else $\mu$m\fi}
\def\kms{\ifmmode {\rm km\,s}^{-1} \else km\,s$^{-1}$\fi}
\def\Hubble{\ifmmode {\rm km\,s}^{-1}\,{\rm Mpc}^{-1}
        \else km\,s$^{-1}$\,Mpc$^{-1}$\fi}
\def\ergsec{\ifmmode {\rm ergs\;s}^{-1} \else ergs s$^{-1}$\fi}
\def\ergscm{\ifmmode {\rm ergs\,s}^{-1}\,{\rm cm}^{-2}
          \else ergs\,s$^{-1}$\,cm$^{-2}$\fi}
\def\ergscmA{\ifmmode {\rm ergs\,s}^{-1}\,{\rm cm}^{-2}\,{\rm \AA}^{-1}
          \else ergs\,s$^{-1}$\,cm$^{-2}$\,\AA$^{-1}$\fi}
\def\ergscmHz{\ifmmode {\rm ergs\,s}^{-1}\,{\rm cm}^{-2}\,{\rm Hz}^{-1}
          \else ergs\,s$^{-1}$\,cm$^{-2}$\,Hz$^{-1}$\fi}
\def\Msun{\ifmmode M_{\odot} \else $M_{\odot}$\fi}
\def\Lsun{\ifmmode L_{\odot} \else $L_{\odot}$\fi}
\def\qo{\ifmmode q_{0} \else $q_{0}$\fi}
\def\Ho{\ifmmode H_{0} \else $H_{0}$\fi}
\def\ho{\ifmmode h_{0} \else $h_{0}$\fi}
\def\qo{\ifmmode q_{0} \else $q_{0}$\fi}
\def\ao{\ifmmode a_{0} \else $a_{0}$\fi}
\def\to{\ifmmode t_{0} \else $t_{0}$\fi}
\def\Halpha{\ifmmode {\rm H}\alpha \else H$\alpha$\fi}
\def\Hbeta{\ifmmode {\rm H}\beta \else H$\beta$\fi}
\def\hb{\ifmmode {\rm H}\beta \else H$\beta$\fi}
\def\Hgamma{\ifmmode {\rm H}\gamma \else H$\gamma$\fi}
\def\Hdelta{\ifmmode {\rm H}\delta \else H$\delta$\fi}
\def\Lya{\ifmmode {\rm Ly}\alpha \else Ly$\alpha$\fi}
\def\Lyb{\ifmmode {\rm Ly}\beta \else Ly$\beta$\fi}
\def\hi{\ifmmode \mbox{{\rm H}\,{\sc i}} \else H\,{\sc i}\fi}

\def\ciii{\ifmmode {\rm C}\,{\sc iii} \else C\,{\sc iii}\fi}
\def\civ{C\,{\sc iv}\,$\lambda1549$}

\def\lnii_ha{L([N\,{\sc ii}])/L(H$_{\alpha}$) }
\def\loiii_hb{L([O\,{\sc iii}])/L(H$_{\beta}$) }

\def\oiii{[O\,{\sc iii}]\,$\lambda5007$}

\def\o5007{[O\,{\sc iii}]\,$\lambda5007$}

\def\ne212m {[Ne\,{\sc ii}]\,$12.8 \mu m$}

\def \Lop{$L_{5100}$}
\def \L5100{$L_{5100}$}
\def \Ledd{$L/L_{\rm Edd}$}

\def  \kms         {\hbox{km s$^{-1}$}}          % kilometers per sec
\def  \ergs        {\hbox{ergs s$^{-1}$}}              % erg/sec
\def  \cmii        {\hbox{cm$^{-2}$}}
\def  \La          {\ifmmode {\rm Ly}\alpha \else Ly$\alpha$\fi}
\def  \Ka          {\ifmmode {\rm K}\alpha \else K$\alpha$\fi}
\def  \Lb          {\ifmmode {\rm L}\beta \else L$\beta$\fi}
\def  \Ha          {\ifmmode {\rm H}\alpha \else H$\alpha$\fi}
\def  \Hb          {\ifmmode {\rm H}\beta \else H$\beta$\fi}
\def  \Pa          {\ifmmode {\rm P}\alpha \else P$\alpha$\fi}
\def  \CIIIb       {\ifmmode {\rm C}\,{\sc iii]}\,\lambda1909
                     \else C\,{\sc iii]}\,$\lambda1909$\fi}
\def  \CIV         {\ifmmode {\rm C}\,{\sc iv}\,\lambda1549
                     \else C\,{\sc iv}\,$\lambda1549$\fi}
\def  \MgII         {\ifmmode {\rm Mg}\,{\sc ii}\,\lambda2798
                     \else Mg\,{\sc ii}\,$\lambda2798$\fi}
\def  \OVI         {\ifmmode {\rm O}\,{\sc vi}\,\lambda1035
x
                     \else O\,{\sc vi}\,$\lambda1035$\fi}

\shorttitle{Black-Hole Mass and accretion rate}
\shortauthors{Netzer}

\begin{document}
%%%%%%%%%%%%%%%%%%%%%%%%%%

\title{Radiation pressure force and black hole mass  determination
in low redshift type-I and type-II active galactic nuclei}

\author{
Hagai Netzer,\altaffilmark{1}
}

\altaffiltext{1} {School of Physics and Astronomy and the Wise
  Observatory, The Raymond and Beverly Sackler Faculty of Exact
  Sciences, Tel-Aviv University, Tel-Aviv 69978, Israel}

\begin{abstract}

The distributions of L(\oiii), black hole (BH) mass and \Ledd\ in two large
samples of type-I and type-II  
active galactic nuclei (AGNs) are compared in order to test the suggestion
that radiation pressure force is affecting the gas velocity in the broad
line region and hence the BH mass determination.
The samples are drawn from the SDSS archive
and are modified to represent the same parent distribution
at $0.1 \leq z \leq 0.2$. BH masses  
in type-I sources are calculated in two different ways,
one using a simple
virial mass assumption and the other by taking into account the 
effect of radiation pressure force on the gas. 
The simple virial mass estimate results in good agreement with the $\sigma_*$-based  
BH  mass and \Ledd\ estimates in type-II sources.
In contrast, there is a clear disagreement in the \Ledd\ distributions when
radiation pressure-based estimates are used. This
indicates that radiation pressure force
is not important in  $0.1 \leq z \leq 0.2$ AGNs with
\Lop=$10^{42.8-44.8}$ \ergs. This has important implications to the physics of
the gas producing the broad emission lines in AGNs, in particular the
existence of extremely
large column density ($\sim 10^{24}$ \cmii) clouds.

\end{abstract}

\section{Introduction}

Black hole (BH) masses in thousands of type-I active galactic nuclei (AGNs) can now be
obtained by combining measurements and understanding of the gas in
the broad line region (BLR) of such sources. 
Information on the gas kinematics in the BLR is obtained
from the study of emission line profiles which leads to the conclusion that, in
many sources, the motion is virialized and completely dominated by the BH gravity 
(e.g. Onken and Peterson 2002). This provides a simple way of estimating the mean Keplerian motion
of the gas e.g. from measured FWHMs of broad emission lines.
Size measurements of the BLR are the result of reverberation mapping (RM)
in a small ($\sim 35$) number of  low redshift ($z < 0.3$) AGNs.  
This provides a scaling relationship between the BLR size and the source luminosity.
The combination of BLR size and mean gas velocity is the basis for the so-called
``virial'' (or ``single epoch'')  
method (Vestergaard and Peterson 2006  and references therein) for obtaining 
BH masses (\MBH) from spectroscopic observations of type-I AGNs.

The accuracy of the virial method depends on various assumptions about the
RM method, the source luminosity 
and the line profiles.
 Detailed discussion of RM and its limitations,  and
listing of the available data, are given in Kaspi et al. (2000, hereafter K00)
and Kaspi et al. (2005; hereafter K05). Improvements and  additions to
the K00 sample are described in Bentz et al. (2008; hereafter B08) where full
account of the host galaxy contribution to the observed flux is included. 
Errors and uncertainties are
discussed in Peterson and Bentz (2006)  and various other publications
and are estimated to be a factor of $\sim2$ on \MBH. This is a combination of the
uncertain host galaxy flux
in low luminosity AGNs, the uncertainties on the
measured time lags, and uncertainties in the conversion of observed
broad line profiles (in this work FWHM(\hb)) to a ``mean gas velocity''.

An important recent development is the suggestion by Marconi et al. 
(2008; hereafter M08) 
that radiation pressure force operating on the
BLR gas can play an important role in determining the gas dynamics. Such
force results in a reduced effective gravity which 
is translated to a larger \MBH\ for a given observed FWHM.
The treatment of radiation pressure depends on the assumed gas distribution
in the BLR. It is straightforward in the case of the ``cloud model'' where the gas
is assumed to be distributed in large column density condensations (clouds) or filaments
(e.g. Netzer 1990 and references therein). It is more complicated 
in other models e.g. the ``locally optimally emitting clouds (LOCs)'' scenario
 (Baldwin et al. 1995; Korista et al. 1997) where
many different components with different density, column density and level of ionization occupy
the same volume of space. The M08 suggestion is not relevant to those models where
the material is assumed to be driven away from a central disk in form of a clumpy wind
(e.g. Everett 2005; Elitzur \& Shlosman 2006). In such cases the virial assumption breaks down and  the line profiles 
 cannot be used in the estimate of \MBH\
In the following I will only consider the cloud model which is the focus of
the M08 paper.

The effect of radiation pressure force is most noticeable in high luminosity sources where
the radiation pressure acceleration can be very large.
This is illustrated in the recent analysis by Marconi et al. (2008b) who applied 
their corrected (i.e. after applying the radiation pressure force term, see below)
mass estimates to the large AGN sample of Shen et al. (2008). A clear signature of the correction
is seen in the distribution of the normalized accretion
rate, \Ledd, which is narrower and hardly ever exceeds 0.1 (Marconi et al. 2008b Fig.\,1).
In contrast, 
the simple virial method results in a substantial fraction of sources  with  \Ledd$\simeq 1$.

This paper proposes a novel method to test the M08 suggestion by comparing two
large samples of type-I and type-II AGNs drawn from the
Sloan Digital Sky Survey (SDSS; York et al. 2000). The assumption that the two
 are drawn from the same parent distribution enables the comparison of BH masses that are
obtained by two independent methods. This provides 
a direct test of the importance of radiation pressure force in accelerating the BLR gas.
In \S2 I describe the two samples and their selection.
\S3 compares the mass and accretion rate distributions  
and contrast them with the M08 suggestion. 
 Finally, \S4 gives a short discussion of the uncertainties and the implications to
higher redshift higher luminosity AGNs.

\section{Black hole mass and accretion rate distributions}

\subsection{Low redshift type-I and type-II AGN samples}

The present work is based on the analysis of two samples of type-I and type-II 
AGNs that cover
the same range in redshift and luminosity. The type-I sample is the one described in
Netzer and Trakhtenbrodt (2007; hereafter NT07). It includes all SDSS Data Release Five (DR5) type-I
AGNs with $z \le 0.75$. The redshift limit is dictated by the need to measure
the broad \hb\ line, the optical continuum luminosity ($\lambda L_{\lambda}$ at
5100\AA; hereafter \Lop) and the \oiii\ narrow emission line in the SDSS spectra.
The extraction, line and continuum fitting,
 and luminosity and \MBH\ determinations  are explained
in NT07.  The values of \Lop\ and \MBH\ 
are very similar to the ones listed in the recent compilation
of  Shen et al. (2008). Radio laud (RL) AGNs are excluded
from the present work (see below) but their inclusion changes nothing in the 
type-I type-II comparison.

The type-II sample is an extension of the SDSS/DR1 sample 
discussed in Kauffmann et al. (2003) and Heckman et al. (2004).
It is based on the DR4 release (Adelman-McCarthy et al., 2006) and 
is publicly available on the MPA site\footnote{www.mpa-garching.mpg.de/SDSS/DR4/}. 
For each source it includes redshift,  stellar velocity
dispersion measured through the 3\arcsec\ SDSS fiber, L(\oiii) 
with and without reddening correction, \loiii_hb, \lnii_ha\ and various other 
properties that are not relevant to the present work. 

The above two samples are different in two major ways:\\
{\bf 1. AGN type:} While the type-I sample includes only broad line AGNs, 
the type-II sample contains both high excitation (Seyfert and QSO)
and low excitation line (LINER) AGNs (starburst galaxies have
been removed using
standard line ratio diagrams; see Kauffmann et al. 2003). 
It is therefore important to identify similar ionization and excitation ranges and to
remove the LINERs from the type-II sample prior to the comparison.
 The separation is done by applying two criteria:
\loiii_hb$>1.4$ (the lowest value in the NT07 sample) and   
\begin{equation}
\log L([N\,II])/L(H_{\alpha}) \leq 
\log L([O\,III])/L(H_{\beta}) - 0.4 \,\, .
%\log \lnii_ha \leq 0.5[\log \loiii_hb -0.4] \,\, .
\label{eq:ration}
\end{equation}
%
%log(\lnii_ha)$\leq$0.5[log(\loiii_hb) -0.4]. Inspection of the remaining Seyferts and QSOs
 The resulting type-II sub-sample is in good agreement with other work of this type (e.g. 
Kauffmann et al. 2003;  Groves et al. 2006) and its mean excitation level is in agreement with the NT07 sample.

{\bf 2. Flux limit and redshift distribution:}
   While the original type-II sample of Kauffmann et al. (2003) and Heckman et al. (2004)
is selected from the flux limited SDSS galaxy sample, the extended DR4 catalog
used here is defined in a different way and includes all sources classified
as galaxies. These are further classified according to the emission line intensities,
in particular \oiii.
This selection reaches
very low L(\oiii) at low redshift.
 The type-I selection is a combination of SDSS colors in a flux limited
($i$=19.1 mag)  sample and the detection
of broad emission lines. 
This results in a noticeable difference between  low
and high redshift sources.
For high redshift type-I sources that are dominated by the AGN continuum,
the flux limit is the determining factor. For low redshifts low luminosity AGNs,
contamination by the host galaxy makes the detection of broad emission lines
more difficult and,
in many cases, becomes the limiting selection factor.
Inspection of the \Lop\ distribution in the NT07 sample
 suggests that the detection of broad emission lines
is the limiting factor for all 
$z \le 0.1$ AGNs. 
This corresponds to \Lop$ \simeq  10^{42.8}$ \ergs\ for the faintest sources
at $z=0.1$.
 For $z \geq 0.15$, the distribution of 
 \Lop\ is consistent with the 
flux limit of the sample. Thus, at $ z \sim 0.1$,
  the type-II sample is deeper
in terms of L(\oiii) and the type-I sample is incomplete.

The distribution of L(\oiii) in type-I AGNs was studied in various
papers including Kauffmann et al (2003), Zakamska et al. (2003) and Netzer et al. (2006). 
The space distribution of all AGNs, based on L(\oiii), is studied in other recent
publications (Reyes et al. 2008 and references therein).
Netzer et al. (2006) show a weak dependence of L(\oiii)/\Lop\ on redshift
and a stronger dependence on \Lop\
within different redshift intervals. For $ z \sim 0.1$, the mean and the 
median values in the NT07 sample are similar and suggest
\Lop/L(\oiii)$\simeq 340$. 
This ratio, which is not corrected for reddening (see below),  is assumed to represent 
all sources in the present work.

To produce two samples that are compatible in redshift and in L(\oiii),
I chose a flux limit for the type-II sources that corresponds to
\Lop$=10^{42.8}$ \ergs\ at $z=0.1$, very close to the flux of the
lowest luminosity type-I
AGNs at this redshift.\footnote{Standard cosmology 
with H$_0$=70 km/sec/Mpc, $\Omega_m=0.3$
and $\Omega_{\Lambda}=0.7$, is used throughout.}
The resulting type-II sample shows a clear lack of sources 
at $z \ge 0.2$; a  behavior that has been
noted in several SDSS-based publications.
Given the deficiency of type-I sources at $z \leq 0.1$,
and the incompleteness of the type-II population at
$z \geq 0.2$, the compromise chosen here is to focus on  
the $0.1 \leq z \leq 0.2$ range. In this range $10^{42.8} \leq$\Lop$ \leq 10^{44.8}$ \ergs. 

Fig.\,1 compares the
L(\oiii) distributions of the two samples in two redshift intervals,
0.1--0.15 and 0.15--0.2. 
There are 4197 type-II and 1331 type-I sources in the chosen
luminosity and redshift ranges and the ratio is close  
to the expected  4:1 (note
that the type-II sources  are from DR4 and the type-I from DR5).
The type-I sources in the histogram are all  radio quiet (RQ) and 
are the same objects discussed in NT07. 
This choise is consistent with the SDSS AGN-detection
algorithms that select radio detected AGNs for spectroscopy regardless of their
color.
Tests show that a distribution that includes also  RL sources is
almost indistinguishable from the one shown here.
All comparisons described below were carried out, separately,
 for the above two redshift bins and
all show the same results. Given this, the rest of the paper addresses the entire 
 $0.1 \leq z \leq 0.2$ range.

\begin{figure}
\plotone{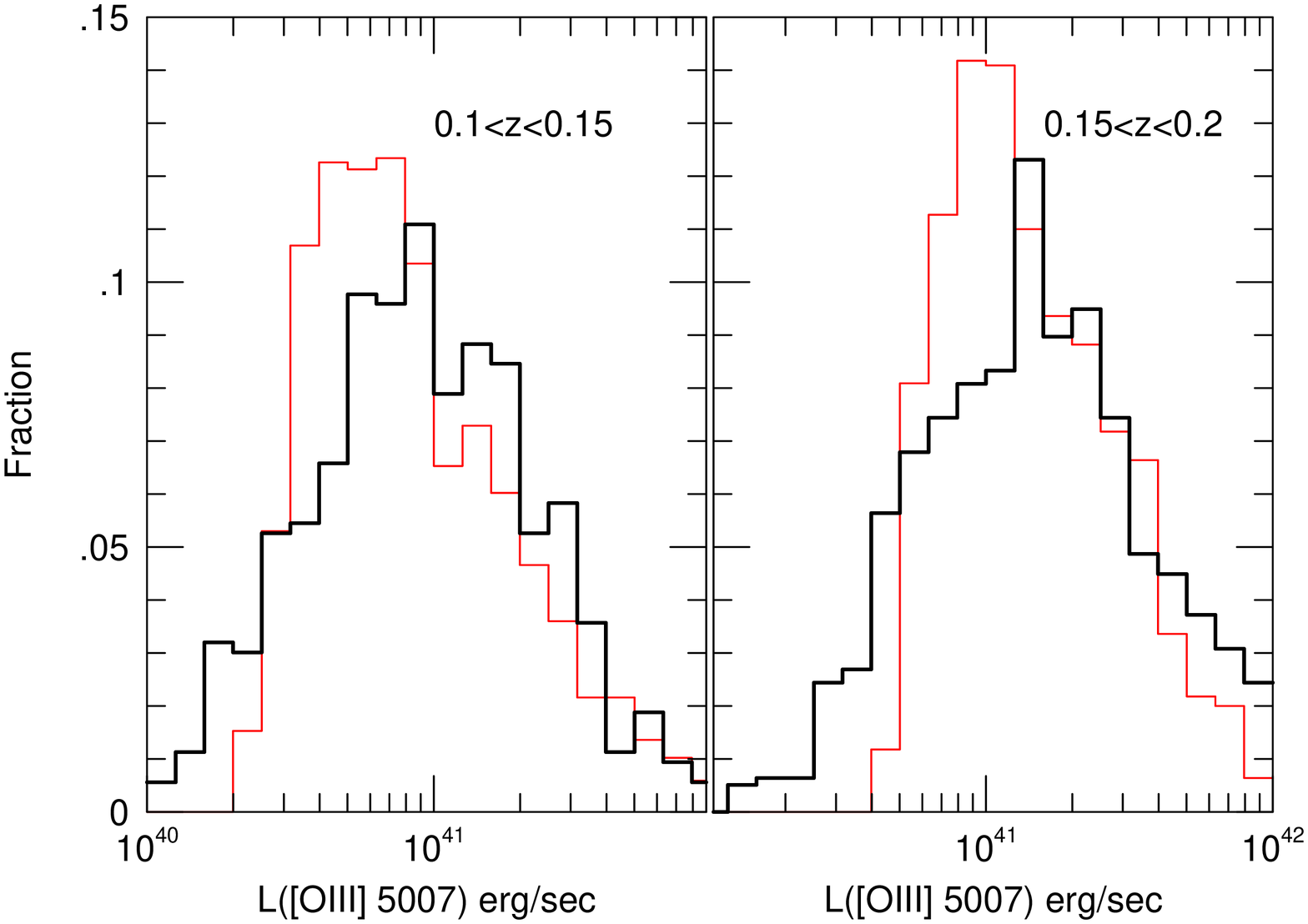}
\caption{L(\oiii) distributions for type-I (black) and type-II (red) AGNs in
two redshift intervals, as marked, using the flux limits defined in the paper.}
\label{fig:Loiii}
\end{figure}

The similarity of the type-I and type-II L(\oiii)  distributions 
(note the shift between the two redshift bins  due to the changing flux limit)
suggests that they are drawn from the same parent 
population.  The somewhat different slope at the low L(\oiii) end is due to
the more complete type-II sample at $z \sim 0.15$ compared with
type-I AGNs that are still affected by host galaxy contamination.
The small deviation on the high L(\oiii) side is either due to the chosen 
flux limit or to reddening of the \oiii\ line. This is discussed in \S4.

\subsection{Mass and accretion rate distributions}

BH masses for the modified type-II sample were obtained by applying the 
Tremaine et al. (2002) expression to the measured $\sigma_*$ (e.g. Heckman et al. 2004).
There are uncertainties in such mass estimates related to the
host type (spheroid dominated, disk dominated,
etc.). However, type-II AGNs reside almost exclusively in massive galaxies 
(Kauffmann et al. 2003) and the above procedure is the only one available for such sources.

 For the type-I sources I used the following  expression,
\begin{equation}
{\rm M_{BH}}= 10^a L_{44}^{0.6} \left[ \frac{FWHM(H_{\beta})}{1000 \,\, {\rm km \, s^{-1}}} \right ] ^2
    \,\,  \Msun  \, \, ,
\label{eq:HN}
\end{equation}
where $L_{44}$=\Lop/$10^{44}$ \ergs\ and $a=6.7$.
The factors in this expression  are adopted from the recent work of B08
that study the {\it intrinsic} R$_{\rm BLR}$ vs. \Lop\ correlation in  
the  K00 sample by using more accurate 
host galaxy subtraction. The subtraction of the host light  affects the slope and the
normalization of the relationship 
which are thus different from the ones given in K05.
The best intrinsic slopes found by B08 are between 0.52 and 0.55,
depending on the fitting procedure with an
uncertainty of order 10\%. The B08 expression cannot be used here since 
the  values of \Lop\ are obtained from SDSS fluxes measured through the 3\arcsec\ fibers 
that include the host galaxy flux. This can be significant in low luminosity
AGNs.

Eq.~\ref{eq:HN} takes the galaxy contribution into account by estimating a 3\arcsec\ host
 flux for each of  the B08 sources using data listed in
that paper. These fluxes were added to the intrinsic B08 AGN fluxes  
and the best R$_{\rm BLR}$-\Lop\ relationship  was found by using a fitting procedure identical to the one
used in that paper. The results are based on fitting  only those RM
sources with \Lop$> 10^{42.5}$ \ergs\ since these are
the ones studied in the present work. The uncertainties on the slope (0.6) and
the normalization ($a$) are about 10\% and their  exact combination is of 
little significance over the limited luminosity range used here.  

M08 suggested that radiation pressure force plays an important
role in changing the velocity of BLR clouds with column density of 
 $\sim 10^{23}$ \cmii\ or smaller. In the cloud model, the column density of BLR clouds  are estimated
 to be of this order because of various considerations
involving the emitted spectrum (Netzer 1990 and references therein). For example, low ionization
lines of FeII and MgII are observed to be very strong yet they cannot originate from the 
highly ionized, illuminated part of the clouds whose column density approaches $10^{22}$ \cmii.
Thus a more extended low ionization part of the clouds is inferred.
There is no clear upper limit to the total column density, only
to the part which is ionized enough to produce emission lines.
The expression adopted for the present calculations is the
one presented in Marconi et al. (2008b),
\begin{equation}
{\rm M_{BH}}= 10^{a_{rad}} L_{44}^{0.5} \left [ \frac{FWHM(H_{\beta})}{1000 \,\,  {\rm km \, s^{-1}}} \right ] ^2
    + 10^b \frac{L_{44}}{N_{23}}  \,\,  \Msun  \, \, ,
\label{eq:M}
\end{equation}
where $N_{23}$ is the hydrogen column density in units of 10$^{23}$ \cmii.
All calculations shown below assume $N_{23}=1$. 

M08 analyzed a small sample of type-I AGNs with both RM-based and
$\sigma_*$-based \MBH. Minimizing the scatter between the two methods  
they derived the following values for the virial and radiation pressure force terms:
$a_{rad}\simeq 6.13$ and $b \simeq 7.7$.
An estimate of $b$ can also be obtained from simple estimates 
of the radiation pressure
force operating on BLR clouds exposed to a standard AGN 
continuum.
This value is in  good agreement with the value obtained from the minimization procedure.
Note that the  difference 
between $a$ (Eq.~\ref{eq:HN}) and $a_{rad}$ (Eq.~\ref{eq:M}) amounts to a factor of 
$\sim 4$ in \MBH.
Most results discussed in this paper
use these values and \S4 addresses the possibility of a larger $a_{rad}$.

The above assumptions and expressions have been  used
to obtain \MBH\ for all sources. There is one
estimate for each of the type-II sources, based on the \MBH-$\sigma_*$
relationship, and two for each of the type-I sources, one that assumes 
Eq.~\ref{eq:HN} and one that assumes Eq.~\ref{eq:M}.
The histograms of all three \MBH\ estimates are shown in Fig.\,2. 

\begin{figure}
\plotone{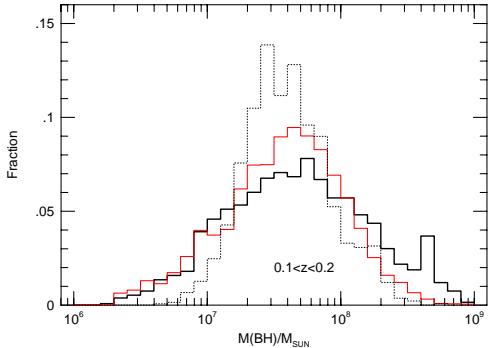}
\caption{\MBH\ distributions of type-I (thick black line) and type-II (thin red
line) $0.1 \leq z \leq 0.2$ AGNs. \MBH\
 estimates that include a radiation pressure force term
(Eq.~\ref{eq:M})
are drawn with a dotted black line}
\label{fig:M}
\end{figure}

I used several statistical tests to compare the histograms.
Since the KS test is not very useful for 
comparing very large samples, I have adopted a practical alternative based on 
a realization procedure. It involves 
choosing, in a random way, a small fraction (2--10\%) of the two 
populations and applying the KS test to the two partial samples.
This was done a large number of times and the range of the resulted
KS-p probability was examined.
The result is a high probability
that all \MBH\ distributions are consistent with each other.
For example, when testing the virial mass estimate in type-I sources 
vs. the $\sigma_*$
mass estimate in type-II sources, 
using  3\% of the objects in each realization, I find 
50\% of the cases to be in the range 
$0.07 \leq p \leq 0.5$, i.e. consistent with the 
assumption of the same parent distribution.
A very similar range of KS-p is obtained when replacing the 
virial assumption with the M08 assumption.
I have also used the Mann-Whitney U-test, in a similar realization
manner, to check for differences in
the medians of the various populations.
The results for the comparison of the \MBH\ distributions are similar
to those of the KS test.

I have computed \Ledd\ for all sources where $L$ is the
bolometric luminosity. Since only L(\oiii) are available
for estimating L in type-II sources, this  
 was  used to obtain L in  in type-I objects. 
For this purpose I assumed a simple bolometric correction factor (BC) to convert \Lop\
to $L$ with the conversion factor given earlier (340) to convert L(\oiii)
to \Lop. The factor BC 
is similar to the one used by Marconi et al. (2004)
and various other recent publications\footnote{Note that some recent papers,
including Shen et al. (2008) and  Hopkins et al. (2007), give larger bolometric corrections. This is
the result of the double counting of the mid-Infrared part of the spectrum
that includes mostly processed radiation.} and the form adopted here is,
\begin{equation}
BC= 9- \log L_{44} \,\, .
\end{equation}
Fig.\,3 shows three \Ledd\ distributions corresponding to the three
\MBH\ distributions of Fig.\,2. Inspection of Fig.\,3 shows
good agreement between
the virial \Ledd\ distribution for type-I sources  and the $\sigma_*$ \Ledd\ 
distribution for type-II
sources. This is also confirmed by the KS  realization tests
 ($0.07 < p < 0.48$ in 50\% of all realizations).
 The distribution of \Ledd\ based on Eq.~\ref{eq:M}
is clearly different showing
a large peak near \Ledd$\sim 0.1$ and almost no source close to \Ledd=1.
For example, the range $ 0.05 \le$\Ledd$ \le 0.2$ contains 30\% of all type-I 
sources assuming Eq.~\ref{eq:HN}
and 74\% of type-I sources assuming Eq.~\ref{eq:M}.
The KS realization tests confirm this result showing 
that 50\% of all realizations produce 
$10^{-4} \leq p \leq 10^{-5}$; i.e extremely small probability for similar
parent distributions.
Unfortunately, the Mann-Whitney test is not very useful in this case since
all distributions have similar mean \Ledd\ (Fig.\,3)
 despite their very different shapes.
\begin{figure}
\plotone{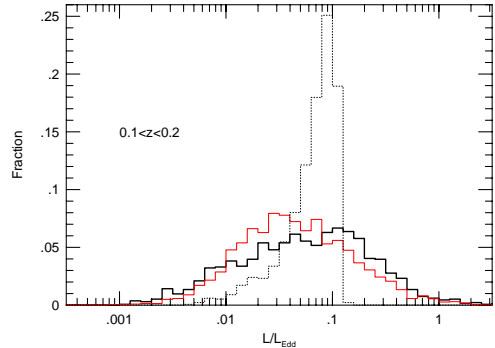}
\caption{\Ledd\ distributions for type-I and type-II sources
(symbols as in Fig.\,2). Note the highly peaked distribution and
the lack of large \Ledd\ sources when radiation pressure force is included.}
\label{fig:Ledd}
\end{figure}

\section{Discussion}

The aim of the present paper is to investigate the mass
and accretion rate distributions in AGNs
hence it is crucial to understand the differences between the 
type-I and type-II samples.
As explained in \S2, the difficulties in defining two identical samples  are
related to the incompleteness of the type-I sample at $z \le 0.1$
and the deviation from a pure flux limited sample of type-II sources at 
$z>0.2$. The chosen redshift interval of 0.1--0.2
is a compromise which is by no means  perfect.
For example, the matching of the two L(\oiii) distributions shown in the
left panel of  Fig.\,1 ($1 \leq z \leq 0.15$ sources) is
improved if the flux limit (\S2) 
is increased by 0.1-0.2 dex. However, there is no obvious physical reason for such
an arbitrary scaling.

Reddening of the \oiii\ line  is perhaps a more plausible  explanation
for the differences seen in Fig.\,1.
This issue has been discussed
extensively in Netzer et al. (2006; see \S3.2)
 where references to earlier works are
given. According to that paper, 
 a comparison of L(\oiii) with L(2--10 keV)
(assumes to
be orientation independent) in type-I and type-II AGNs indicates more extinction in type-II
sources. The typical difference is of order 0.2--0.3 dex
but the number is highly uncertain since it is derived from a mixture of
sources that include optically selected and X-ray selected AGNs (note that
the above factor represents the {\it difference} in extinction, not the extinction
itself which is directly measured in most type-II samples
and is typically larger). 
More support for differences in reddening  can
be found in the comparison of the L(\oiii)-based luminosity functions of 
type-I and type-II AGNs (e.g. Reyes et al. 2008) and in the comparison of optical and
mid-IR emission lines 
(Mel{\'e}ndez et al.2008).
Applying such a correction to the type-II sample used here  brings the
L(\oiii) distributions shown in Fig.\,1 into  a better agreement
with the equivalent type-I distribution.
The correction 
does not affect the type-II BH mass distribution
 (which is based on 
$\sigma_*$) but increase the deduced \Ledd\ for type-II AGNs. 
This improves the (already good)
agreement shown in Fig.\,3.

I have also investigated the possibility that most of the
differences between the two methods used to derive \MBH\ in type-I sources stem from the
difference between $a$ and $a_{rad}$. This is in accord with a new work by Marconi
and collaborators who are studying this idea in samples of higher redshift type-I sources 
(A. Marconi, private communication).
To test this idea I chose a larger value, $a_{rad} =6.5$, in Eq.~\ref{eq:M}  and
calculated new \MBH\ and \Ledd\ distributions. The result is a somewhat better agreement with the type-II
\MBH\ distribution
but the \Ledd\ distribution changed only slightly and shows
the same typical deficiency of \Ledd$\sim 1$ sources shown in Fig.\,3.
To illustrate this more clearly, I show in Fig.\,4 the two dimensional distributions
of \MBH\ vs. \Ledd\ in all three cases: one for type-II sources (red points) and two for the
type-I AGNs, the virial method (black points) and the M08 expression with $a_{rad}=6.5$ (blue points).
The very clear deviations of the Eq.~\ref{eq:M}-based estimates are evident.

\begin{figure}
\plotone{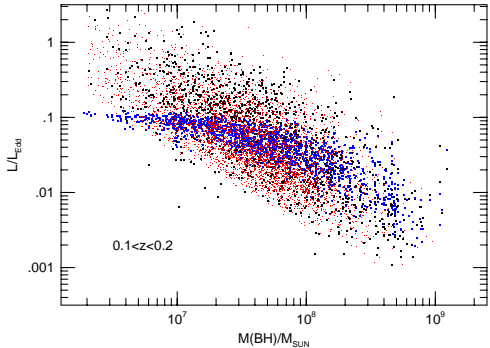}
\caption{\Ledd\ vs. \MBH\ for type-II sources (red points) and for two assumptions about
type-I sources: simple virial assumption (black points) and modified estimates
(Eq.~\ref{eq:M}) that include
 radiation pressure force term with $a_{rad}=6.5$ (blue points). Note the reduced ranges in \MBH\ and
\Ledd\ and the very different distribution compared with type-II sources,
 when the modified estimate is used.
 }
\label{fig:M_Ledd}
\end{figure}

 Finally, I have also experimented
with changing the slope in Eq.~\ref{eq:M} to
0.6, similar to the one used in Eq.~\ref{eq:HN}. This had negligible effect 
on the results. 
There are other potential complications, e.g. the possibility that the L(\oiii)/\Lop\
ratio is itself a function of \Ledd, the suggestion that the bolometric correction factor, 
BC, depends on \Ledd, and more. The data used here do not support the first and there is no
information to test the latter. 

The main result of the present work is the significantly 
different \Ledd\ distribution
of low luminosity low redshift 
type-I AGNs compared with type-II sources under the assumption that radiation pressure force
plays an important role in affecting the BLR gas velocity.
The differences almost completely
disappear when a ``standard'' virial mass estimate (Eq.~\ref{eq:HN}) is
used. Eq.~\ref{eq:M} shows that the radiation pressure force term is negligible if
the column densities
of the \hb\ producing BLR clouds are significantly larger than 10$^{23}$ \cmii,  perhaps
as large as $10^{24}$ \cmii. Radiation 
 pressure force bounds to have some 
effect on optically thick gas thus the  results shown here 
can be viewed as an indication for large column density BLR clouds. As explained, all this relates 
only to the cloud model of
the BLR.
 
The existence of extremely large column density clouds can change some aspects of present
AGN models.
For example, it limits the possibility of escaping BLR gas and increases the chance of
eventual accretion onto the central BH. It also suggests a large column of neutral gas at the
back of such clouds and a more hospitable environment for molecules and dust
 formation and survival. The
physical scale of the clouds must be larger too compared with previous estimates with possible implications
for cloud-cloud collisions.
Regarding high redshift sources, the Marconi et al.  (2008b) mass calculations
is based on the observation of the \civ\ line. Since  
the column densities of \hb\ producing and \civ\ producing clouds are not
necessarily the same (e.g. Kaspi and Netzer 1999)  this would
complicate  the comparison with low redshift samples.

Finally, the present work applies to AGNs with $10^{42.8} \leq$\Lop$\leq 10^{44.8}$ \ergs\
in the 0.1--0.2 redshift interval. I have also tested the \Ledd\ distribution
in somewhat higher luminosity higher redshift ($\sim 0.25$) type-II sources from the 
DR4 archive. I found several
cases with \Ledd$\sim 1$, beyond the limit obtained by using Eq.~\ref{eq:M}.
As for a similar test for higher redshift, more
luminous sources, the situation is less clear. First it is hard to find overlapping
type-I and type-II samples where a similar analysis can be applied. 
Moreover, the applicability of the \MBH-$\sigma_*$ relationship
 has only been demonstrated at low redshift. 
The current understanding of the co-evolution of massive BHs and their hosts at high redshift 
does not allow such a test at the present time.

\acknowledgments I am grateful to Alessandro Marconi, Benny Trakhtenbrot and 
Guinevere Kauffmann for useful discussions.
Funding for this work has been provided by the Israel Science
Foundation grant 364/07. 

Funding for the SDSS and SDSS-II has been provided by the Alfred P. Sloan 
Foundation, the Participating Institutions, the National Science Foundation, 
the U.S. Department of Energy, the National Aeronautics and Space 
Administration, the Japanese Monbukagakusho, the Max Planck Society, 
and the Higher Education Funding Council for England. 
The SDSS Web Site is http://www.sdss.org/.

    The SDSS is managed by the Astrophysical Research Consortium for the 
Participating Institutions. The Participating Institutions are the 
American Museum of Natural History, Astrophysical Institute Potsdam, 
University of Basel, University of Cambridge, Case Western Reserve University,
 University of Chicago, Drexel University, Fermilab, the Institute for 
Advanced Study, the Japan Participation Group, Johns Hopkins University, 
the Joint Institute for Nuclear Astrophysics, the Kavli Institute for 
Particle Astrophysics and Cosmology, the Korean Scientist Group, the Chinese 
Academy of Sciences (LAMOST), Los Alamos National Laboratory, the 
Max-Planck-Institute for Astronomy (MPIA), the Max-Planck-Institute for 
Astrophysics (MPA), New Mexico State University, Ohio State University, 
University of Pittsburgh, University of Portsmouth, Princeton University, 
the United States Naval Observatory, and the University of Washington.

\end{document}